\documentstyle [12pt]{article}
\date{}                   
\everymath{\displaystyle}
\newcommand{\mb}[1]{\mbox{\boldmath$\bf#1$}}

\begin{document}           
\begin{center}
{\large{\bf    Electric Current Multipole Moments
                  in Classical Electrodynamics}
\medskip\medskip

{\large A.J. Silenko}}
\medskip\medskip

{\it Institute of Nuclear Problems,\\
 Belarusian State University, Minsk 220080, Belarus}
\end{center}
\medskip\medskip

{\it The general theory for electric current multipoles appearing
at the motion of magnetic dipoles and change in these values or
orientation has been suggested. Static multipoles, including an
anapole, have been studied in detail.}


\begin{center}
{\bf  1. Introduction }
\end{center}

Electric current multipoles (ECM) occur at the
motion of conductors carrying a current or at the variation of the current
strength in conductors. While possessing a number of typical
multipole properties, they have some interesting distinctions. Typically
electric charge multipoles occur due to a particular spatial distribution
of charge. ECM, as distinct from charge multipoles, are
generated only by conductors carrying a current. Noncompensated charges are
absent in the rest system of any conductor, and the charge density is
equal to zero. There are two types of ECM. In laboratory system for the
ECM occurring at the motion of conductors carrying a current the charge
density is nonzero. Its appearance is due to the transformation of the
charge and current densities as components of a
four-vector. Just the charge density occurring at the motion of
conductors determines electric multipole moments of the system. For this
type of multipoles, electric current dipoles are known long ago
[1], the classical theory of electric current quadrupoles is given in
[2,3], their quantum theory is given in [4,5]. The existence of the
second type of ECM has been suggested by  Miller [6].  Multipoles of
this type appear at the change of a current strenght in a conductors or
at the change of the spatial orientation of conductors carrying a current
(or magnetic multipoles).  The introduction of a magnetic current is a
convenient approach for description of these multipoles [6]. It is
obvious that electric multipoles  of the both types appear at the
arbitrary motion of conductors  carrying a current. This is a general
case and will be analyzed in this work in terms of the classical
electrodynamics.

The investigation of ECM is of a great practical importance. The
electric current dipole moment occurring at the motion of a particle
having a magnetic dipole moment determines the interaction of the
particle's spin with an electrostatic field. The electric
current quadrupole moment (ECQM) appearing at the orbital motion of
nucleons can make a distinct contribution to the total quadrupole
moments of nuclei [3--5]. As shown in [7], the moving anapole
interacts with an inhomogeneous electric field. The investigation
of this phenomenon is of a great interest because it concerns the
non-contact interaction of the anapole with the
static external field. As will be shown below, the interaction of
an anapole with the inhomogeneous electric field is due to the possession
of ECQM by a moving anapole.

\begin{center}
{\bf  2. The Field of Electric Current Multipoles}
\end{center}

    The electric current multipole moments of microscopic objects are
most interesting for the investigation. For these objects, it is
     sufficient to investigate the motion of point magnetic dipoles
     rather than that of conductors of finite sizes, carrying
     a current. In case of microscopic objects, when the sizes of
     conductors carrying a current should be taken into account, the
     formulae obtained for magnetic dipoles can usually be applied with
substitution a set of point magnetic dipoles for conductors carrying a
     current (the magnetic sheets model [8]).

     We obtain the expression for the electric field generated by point
     magnetic dipoles. The strength for this field is defined by the
     formula:
\begin{equation} \mb E=-\nabla\phi-\frac1c\frac{\partial\mb A}{\partial
t}.\label{eq1} \end{equation}

Let $\mb r$ be a radius-vector of a point
     where a point magnetic dipole is found, and let us define the
     strength of the field at a remote point having radius-vector $\mb R$,
where $|\mb R|\gg|\mb r|$. Since we analyze the systems that
     can radiate electromagnetic waves, we emphasize that $|\mb
R|\ll\lambda$, where $\lambda$ is a characteristic wavelength  of the
radiation. In the reference system where a dipole  rests,  a scalar
     potential $\phi'=0$, and in the laboratory system \begin{equation}
     \phi '=(\mb v\cdot\mb A)/c\equiv(\dot{\mb r}\cdot\mb A)/c,
\label{eq2} \end{equation}
 where $\mb v$ is the velocity of the translational motion of a magnetic
     dipole. Since
$$   \mb A=\frac{\mb \mu\times\mb r'}{r'^3}, ~~~ \mb r'=\mb R-\mb r,$$
then by expansion of the quantity $\mb A$ into a series we obtain:
\begin{equation}
\mb A=-[\mb \mu\!\times\!\nabla]\frac{1}{R}+(\mb r\!\cdot\!\nabla)[\mb \mu
\!\times\!\nabla]\frac{1}{R}-\frac{1}{2}(\mb r\!\cdot\!\nabla)(\mb r\!\cdot
\!\nabla)[\mb\mu\!\times\!\nabla]\frac{1}{R}+\dots+(-1)^{k+1}\frac{1}{k!}
(\mb r\!\cdot\!\nabla)^k[\mb \mu\!\times\!\nabla]\frac{1}{R}+\dots,
\label{eq3} \end{equation}
where $(\mb r\cdot\nabla)^k$ is a product k of cofactors $(\mb r\cdot
\nabla)$, and $[\mb\mu\times\nabla]f(R) \equiv [\mb\mu\times\nabla f(R)]$.
Taking the identity $(\mb v\cdot[\mb\mu\times\nabla])=([\mb
v\times\mb\mu]\cdot\nabla)$
into account, the electric field strength could be expressed as:
\begin{equation}
\mb E=\mb E^{(1)} +\mb E^{(2)} +\mb E^{(3)},
\label{eq4} \end{equation}
\begin{equation} \begin{array}{c}
\mb E^{(1)}=-\nabla\phi,~~~ \phi=\phi^{(1)} +\phi^{(2)} +\dots,~~~
\phi^{(1)}=-(\mb d\cdot\nabla)\frac1R,\\ \phi^{(2)}=-(\mb r\cdot\nabla)(\mb
d\cdot\nabla)\frac1R,\dots,~~~ \phi^{(k)}=(-1)^k\frac{1}{(k-1)!}(\mb
r\cdot\nabla)^{k-1}(\mb d\cdot\nabla)\frac1R,   \end{array} \label{eq5}
     \end{equation}
\begin{equation} \begin{array}{c}
\mb E^{(2)}=\nabla\times\mb G,~~~ \mb G=\mb G^{(0)}+\mb G^{(1)}+\mb
G^{(2)}+\dots,~~~\mb G^{(0)}=-\frac{\dot{\mb\mu}}{cR},~~~
\mb G^{(1)}=\frac{1}{c}(\mb r\cdot\nabla)\frac{\dot{\mb\mu}}{R},\\
\mb G^{(2)}=-\frac{1}{2c}(\mb r\cdot\nabla)(\mb r\cdot\nabla)\frac{\dot
{\mb\mu}}{R},\dots,~~~ \mb G^{(k)}=(-1)^{k+1}\frac{1}{ck!}(\mb r\cdot
\nabla)^k\frac{\dot{\mb\mu}}{R},
 \end{array} \label{eq6} \end{equation}
\begin{equation} \begin{array}{c}
\mb E^{(3)}=\nabla\times\mb N,~~~ \mb N=\mb N^{(1)} +\mb N^{(2)} +\dots,~~~
\mb N^{(1)}=\frac1c(\mb v\cdot\nabla)\frac{\mb\mu}{R},\\ \mb N^{(2)}=-
\frac1c(\mb r\cdot\nabla)(\mb v\cdot\nabla)\frac{\mb\mu}{R},\dots, ~~~
\mb N^{(k)}=(-1)^{k+1}\frac{1}{c(k-1)!}(\mb r\cdot\nabla)^{k-1}(\mb v\cdot
\nabla)\frac{\mb\mu}{R},
\end{array} \label{eq7} \end{equation}
where $\mb d=[\mb v\times\mb\mu]/c$. Formulae (4)--(7) describe an electric
field generated both at the translational motion of a magnetic
dipole moment and at the change of its value or orientation. Indices for
$\phi,\mb G,\mb N$ define the rank of multipoles.  The quantity
$-\frac1c\frac{\partial\mb A}{\partial t}$ in formula (1) corresponds to
the sum $\mb E^{(2)} +\mb E^{(3)}$. The potential $\phi^{(1)}$ coincides
with a potential generated by a charge dipole moment $\mb d$, and
formula (5) is the expansion in terms of multipoles. By
separating quadrupole and contact interactions in the potential
$\phi^{(2)}$, the latter can be expressed as:  \begin{equation}
\phi^{(2)}=\frac16Q_{ij}\frac{\partial^2}{\partial X_i\partial X_j}\left(
\frac1R\right)+\frac16\tau\frac{\partial^2}{\partial X_i^2}\left(
\frac1R\right),
\label{eq8} \end{equation}
\begin{equation}
Q_{ij}=3x_id_j+3d_ix_j-2\delta_{ij}x_kd_k, ~~~\tau=2x_kd_k.
\label{eq9} \end{equation}

As is known, the values $\mb d, Q_{ij}$ and $\tau$ represent a dipole
moment, a quadrupole moment tensor, and a mean square  of a charge
     radius, respectively.  For the system of charges
\begin{equation}
\mb d=\!\int\!\rho\mb rdV, ~~~Q_{ij}=\!\int\!\rho(3x_ix_j-\delta_{ij}r^2)dV.
\label{eq10} \end{equation}

The fields $\mb E^{(2)}$ and $\mb E^{(2)}$ are vortex. For immobile magnetic
dipoles $(\mb v=0)~\mb E^{(1)}=\mb E^{(3)}=0$, and their electric field is
the field of ECM described in [6].
The value $\dot{\mb\mu}$ could be nonzero due to the change both
of the magnetic moment value and direction of its orientation.  The
     characteristic feature of this type of ECQM is the possibility of
their description through the introduction of magnetic current of
     density $\mb j^{(m)}=\sum_{i}\dot{\mb\mu}_i \delta(\mb r-\mb
     r'_i)$.  This allows to use
     mathematical tools of electrodynamics for magnetic
charges [9].

    Formula (6) gives the expansion of pseudovector potential $\mb G$
    in terms of degrees $r/R$.  However, in the general case this
     formula does not allow to pass to such expansion in terms of
     multipoles typical for usual electric
     currents.  In particular, the relation typical
    of the magnetic dipoles $$(\mb r\cdot\nabla)\frac{\mb j}
{R}=\frac12\bigl[[\mb r\times\mb j]\times\nabla\bigr]\frac{1}{R},$$
is not fulfilled for
     dipole term $\mb G^{(1)}$ in (6):
$$(\mb r\cdot\nabla)\frac{\dot{\mb\mu}}{R}=\dot{\mb\mu}(\mb r\cdot\nabla)
\frac{1}{R}\neq\frac12\bigl[[\mb r\times\dot{\mb\mu}]\times\nabla\bigr]
\frac{1}{R}.$$

This is due to the fact that the value $\dot{\mb\mu}=\int\mb j^{(m)} dV$
may not be given in the form of  $\dot{\mb\mu}=\int\rho^{(m)}\mb v dV$,
similar to the expression $\mb j=\rho\mb v$.  However, the appropriate
ECM appear in  case of a particular configuration of magnetic currents,
repeating the configuration of electric currents, generating different
magnetic multipoles. The formulae for ECM in this case
coincide with the formulae for magnetic multipoles at the substitution
of $\mb E\!\rightarrow\! -\mb H,~\mb G\!\rightarrow\! -\mb A,~\mb
j^{(m)}\!\rightarrow\!\mb j$. In particular, at the expansion in
terms of multipoles, electric anapole (toroid dipole)
moments appear in a natural way.  Their existence has been suggested
     in [10,11].  The ECM field, with account of terms of the
second order for $r$, is described by the expression:  \begin{equation}
\mb E\!=\!\nabla\!\times\!\mb G,~~
G_i\!=-e_{ijk}d_j'\frac{\partial}{\partial
X_k}\!\left(\frac1R\right)+\left[-\frac16e_{ijn}Q_{nk}+\frac{1}{4\pi}
(\delta_{ik}a_j'-\delta_{jk}a_i')\right]\frac{\partial^2}{\partial
X_j\partial X_k}\!\left( \frac1R\right),
\label{eq11} \end{equation}
where $\mb d, Q_{nk}'$  and $\mb a'$ are the electric current dipole,
quadrupole and anapole moments, respectively:
\begin{equation}
\mb d=-\frac{1}{2c}[\mb r\times\dot{\mb\mu}],
    ~~~Q_{nk}'=\frac{1}{c}(e_{nlm}x_k+e_{klm}x_n)\dot{\mu}_lx_m, ~~~
\mb a'=\frac{\pi}{c}\dot{\mb\mu}r^2.
\label{eq12} \end{equation}

     It is interesting to compare (11),(12) with the corresponding
    formulae for magnetic multipoles [12]:
\begin{equation} \begin{array}{c}
\mb H\!=\!\nabla\!\times\!\mb A,~ A_i\!=\!-e_{ijk}\mu_j\frac{\partial}
{\partial X_k}\!\left(\frac1R\right)\!+\!\left[-\frac16e_{ijn}M_{nk}\!+
\!\frac{1}{4\pi}(\delta_{ik}a_j\!-\!\delta_{jk}a_i)\right]\frac
{\partial^2}{\partial X_j\partial X_k}\!\left( \frac1R\right),\\
\mb\mu\!=\!\frac{1}{2c}\!\int\![\mb r
\times\mb j]dV, ~~~M_{nk}=\frac{1}{c}\!\int\!(e_{nlm}
x_k+e_{klm}x_n)j_lx_mdV, ~~~\mb a\!=\!-\frac{\pi}{c}\!\int\!\mb jr^2dV,
\end{array} \label{eq13} \end{equation}
where $\mb\mu, M_{nk}$ and $\mb a$ are the magnetic dipole, quadrupole
and anapole moments.

The common property of ECM is the coincidence of their fields at great
distances with the fields of corresponding charge multipoles.

Note the principle difference between the two types of ECM. The
    existence of ECM, appearing at the translational motion of
    conductors carrying a current, is related to the appearance of a
    nonzero density of charges:
\begin{equation}
\rho=\frac{\mb j_0\cdot\mb v'}{c^2(1-v'^2/c^2)^{1/2}}=\frac{\mb
j\cdot\mb v'}{c^2},
\label{eq14} \end{equation}
where $\mb v'$ is the velocity of a conductor, $\mb j_0$ is the current
density in the rest system for a conductor carrying a current, $\mb j$ is
the current density in the laboratory system, being a sum of conduction
    and convection currents, $\rho\mb v'$. The appearing charges are
real, since the relation $\nabla\mb E=4\pi\mb\rho$ is satisfied.
Note that the summed-up charge equals zero due to the law of charge
conservation, at arbitrary motion of a conductor carrying a
current in any reference system. The charge density equals zero in the
entire space for ECM appearing at the change of value or direction
of magnetic moments.  These moments occur due to nonzero value
$\frac{1}{2\pi}\frac{\partial \mb H}{\partial t}$, which might be
called the magnetic displacement current by analogy with electric
displacement current. The nonzero density of effective magnetic
charges, defined by the formula similar to formula (14), occurs at
the motion of the ECM of this type. However, magnetic charges and
currents are just effective, and not real, since ECM are described by
the typical Maxwell equations valid for
$\rho^{(m)}=0,~\mb j^{(m)}=0.$

   An essential distinction between the multipoles investigated in
[1--5] and Miller multipoles [6] is related to the type of Lagrangian
and Hamiltonian of the interaction. The field of ECM of the first type
is caused by appearance of a nonzero charge density, and the scalar
potential of this field is also nonzero. Therefore, external charges
interact in the same way with the fields of charge and current
electric moments, and in the both cases the Lagrangian of the
interaction is equal ${\cal L}_{int}=-e\phi$ ($e$ is an external
charge). The situation is different for the multipoles of second
type (Miller multipoles) having unique electrodynamic properties.
The scalar potential of their field is equal to zero, and the Lagrangian
of the interaction of an external charge with the field of a
multipole also equals zero for a charge at rest; ${\cal
L}_{int}=-\frac{e}{c}(\mb v\cdot\mb A)\rightarrow 0$, when the velocity
of the charge $\mb v\rightarrow 0$. However, despite this
circumstance the charge and the Miller multipoles effectively
interact. The force equal to $\mb F=d\mb p/dt=e\mb E$
acts on the charge and affects the
kinetic momentum $\mb p$ of the charge, i.e. causes its motion.
The expressions for ${\cal L}_{int}$ and $\mb F$ agree because  the
constancy of the generalized momentum $\mb P=\partial {\cal L}/\partial
t=\mb p+e\mb A/c=const$ follows from the Lagrange equation:
$$\frac{d}{dt}\frac{\partial {\cal L}}{\partial\mb v}= \frac{d\mb
P}{dt}=\frac{\partial {\cal L}}{\partial\mb v} $$ and the kinetic
momentum $\mb p=m\mb v/\sqrt{1-v^2/c^2}$ varies due to the variation of
the vector potential $\mb A$. So, though the Lagrangian and Hamiltonian
of the interaction of the external resting charge with the Miller
multipole are equal to zero, but their derivatives are nonzero and
this charge begin motion in the Miller multipole field in the same way
as it would move in the field of the corresponding charge multipole. The
motion of the charge in the field of ECM of the both types
outwardly does not differ from its motion in the charge multipole field,
with the moment of the same value.

\begin{center}
{\bf  3. The Interaction of Electric Current Multipoles with the
External Electric Field} \end{center}

   The two types of ECM differ by the character
of their interaction with the external electric field. The Lagrangian
and Hamiltonian of the interaction of resting multipoles of the second
type (Miller multipoles) with the electric field are equal to zero,
because these multipoles have no electric charge. However, here we also
observe the unique properties of the\-se multipoles. Under the influence
of the electric field they begin to move, and their mo\-ti\-on
outwardly has no distinction from the motion of the corresponding charge
multipoles [6].

  The interaction of multipoles of the first type with the electric
fields also has some interesting effects. The charge density, occurring
at the motion of conductors carrying a current and described by
formula (14), defines the type of the Lagrangian of the interaction:
\begin{equation}{\cal L}_{int}=-d_i\frac{\partial\phi }{\partial X_i}
-\frac16Q_{ij}\frac{\partial^2\phi }{\partial X_i\partial X_j}
-\frac16\tau  \frac{\partial^2\phi }{\partial X_i^2},
\label{eq15} \end{equation}
where $X_i$ is  the coordinate  of the center of the current system, and
the values $\mb d,Q_{ij},\tau$ are given by formulae (10) where $\rho$
is defined by formula (14) (with the substitution of $\mb v$ for $\mb
v'$). Note that $\partial^2\phi/\partial
X_i^2\equiv\Delta\phi=-4\pi\rho_{ext}(0)$ where $\rho_{ext}$ is the
external charge density.  In this case the distance to the current
element $\mb r'=\mb R+\mb r$ and $\mb v'\equiv\dot{\mb r'}=\mb V+\mb v$,
where $\mb v\equiv\dot{\mb r},\mb V\equiv\dot{\mb R}$. By taking $\mb
v$ beyond the symbol of the integral we find:
\begin{equation}\begin{array}{c}
d_i=\frac{V_k}{c^2}\int j_kx_idV+\frac{1}{c^2}\int j_kv_kx_idV=d_i^{(1)}+
d_i^{(2)},\\
Q_{ij}=\frac{V_k}{c^2}\int j_k(3x_ix_j-\delta_{ij}r^2)dV+\int j_kv_k
(3x_ix_j-\delta_{ij}r^2)dV=Q_{ij}^{(1)}+Q_{ij}^{(2)},\\
\tau=\frac{V_k}{c^2}\int j_kr^2dV+\int j_kv_kr^2dV=\tau^{(1)}+\tau^{(2)}.
\end{array} \label{eq16} \end{equation}

The first terms in (16) define ECM appearing only at the
motion of magnetic multipoles with the velocity $\mb V$, the second
terms describe ECM differing from zero in the rest system of the
particle. Let us consider at first the first terms. We transform them,
using the definitions of magnetic multipole moments (13). And for the
anapole moment the following expression (see [13]) is equivalent to (13):
\begin{equation}
\mb a=\frac{2\pi}{c}\int(\mb j\cdot\mb r)\mb rdV.
\label{eq17} \end{equation}
It follows from (13),(17) that the formula for the anapole moment can
be also given in the form:
\begin{equation}
\mb a=\frac{2\pi}{3c}\int[\mb r\times[\mb r\times\mb j]]dV.
\label{eq18} \end{equation}

Allowing for that the mean of the time derivative of the value varying in
certain limits is equal to zero, and $j_k=\rho v_k\equiv\rho\dot x_k$
we obtain:
$$ \begin{array}{c}
\langle\frac{d}{dt}(\rho
x_kx_ix_j)\rangle=\langle\dot{\rho} x_kx_ix_j)\rangle +\langle
j_kx_ix_j)\rangle+\langle x_kj_ix_j)\rangle+\langle
x_kx_ij_j)\rangle=0,\\ \langle\frac{d}{dt}(\rho x_kx_i)\rangle=
\langle\dot{\rho} x_kx_i)\rangle +\langle
j_kx_i)\rangle+\langle x_kj_i)\rangle=0.     \end{array}   $$

For stationary currents $\rho=0$. By dropping angular brackets and allowing
for the symmetry by the indexes $i,j$, we find:
$$ j_kx_ix_j\!=\!\frac23(j_kx_ix_j\!-\!x_kj_ix_j)\!=\!\frac23e_{ikl}[\mb r
\!\times\!\mb j]_lx_j\!=\!\frac13e_{ikl}\left([\mb r\!\times\!\mb j]_lx_j\!
+\![\mb r\!\times\!\mb j]_jx_l\!+\![\mb r\!\times\!\mb j]_lx_j\!-\![\mb r\!
\times\!\mb j]_jx_l\right).$$
Hence,
$$ \frac1c\left\langle\int j_kx_ix_j dV\right\rangle=\frac13e_{ikl}M_{lj}+
\frac{1}{2\pi}\left(\delta_{kj}a_i-\delta_{ij}a_k\right). $$
 Similarly we can obtain:
$$ \frac1c\left\langle\int j_kx_i dV\right\rangle=e_{ikl}\mu_{l}. $$

In view of the symmetry by the indices $i,j$ it is easily
found that
\begin{equation} \begin{array}{c}
\mb d^{(1)}\!=\!\frac{1}{c}[\mb V\!\times\!\mb\mu], ~~Q^{(1)}_{ij}\!=\!\frac
{1}{2c}\left(
e_{ikl}V_kM_{lj}+e_{jkl}V_kM_{li}\right)\!+\!\frac{1}{4\pi c}\left(3a_iV_{j}
\!+\!3a_jV_i\!-\!2\delta_{ij}\mb a\!\cdot\!\mb V\right), \\
\tau^{(1)}=-\frac{1}{\pi c}\mb a\!\cdot\!\mb V.
\end{array} \label{eq19} \end{equation}

Formulae (19) for ECM caused by the motion of magnetic dipoles and
magnetic quad\-ru\-po\-les have been obtained in [1] and [2,3],
respectively. Formulae (19) show as well that ECQM also appear as a
result of the anapole motion. The interaction of the moving anapole with
an electrostatic field is a very important nontrivial effect first
found by Afanasiev [7]. In view of the fact that in Ref. [7] a more
particular case of the anapole formed of point magnetic dipoles was
considered, the formulae there obtained agree with (19). We remind that
the field of the anapole has no effect on the
motion of external charges because its strength is equal to zero ($\mb
E\!=\!\mb H\!=\!0$). At the same time, the distinction of the vector
potential from zero leads to the Aharonov-Bomh effect [13]. As is known,
the anapole interacts with an external current at contact, and with an
alternate electric field at distance [14].

   When transferring to the quantum mechanical description, the anapole
moment of a particle should be expressed through the pseudovector $\mb
I$ of the particle spin. According to the formula $\mb a=\frac{a}{I}\mb
I$ [12] the anapole moment also becomes a pseudovector, and the
interaction of the anapole with the electric field is $P$-odd.

   The appearance of ECQM for the moving anapole and its interaction
with the electrostatic field may lead to the observation and measurement
of the electron anapole moment in the experiments with atoms. The
$P$-odd interaction of the electron anapole moment with the
electrostatic field of a nucleus is proportional to the nucleus charge,
and the presence of this interaction can be observed in the precision
experiments with heavy atoms, similar to the experiment where the
anapole moment of $^{133}$Cs was first detected [15].

   As seen from formulae (19), the moving anapole interacts at contact
with external charges. This interaction is proportional to the
value $\tau$. It is also $P$-odd and, in particular, contributes to the
$P$-nonconservation in atoms.

   The second terms in (16) describe nonzero ECM for the particles at
rest. In the absence of the $T$-invariance violation the dipole moments
(including the
electric current moments) of atoms and nuclei are equal to zero. The
ECQM of atoms and nuclei are nonzero, and for nuclei, as shown in
[3--5], they are not small. Their contribution to the total quadrupole
moments of some nuclei comprises the value of a few percent. The
evaluations show that for the nucleus $^{133}$Cs the charge /
current quadrupole moments are of the same order of magnitude [5]. We
remind that the sum of the charge and current quadrupole moments
is determined experimentally. The formulae for $Q_{ij}^{(2)}$ and
$\tau^{(2)}$ are more convenient to be written for point magnetic
dipoles. This expression agrees with the real physical pattern of
this phenomenon for atoms and nuclei. For this purpose the
replacement of $\mb r\rightarrow\mb r+\mb r''$ where $|\mb r''|\ll|\mb
r|$ is sufficient. Then the vector $\mb r$ describes the position
of point magnetic dipoles, $\mb v\equiv\dot{\mb r}$, and the small vector
$\mb r''$ describes the position of current elements, constituting
these dipoles, and $\mb j=\rho\dot{\mb r}''$. The integrals in (16) are
equal to:
\begin{equation}
\frac{1}{c^2}\int j_kv_kx_idV=\frac{1}{c}\sum e_{ikl}v_k\mu_l, ~~~
\frac{1}{c^2}\int j_kv_kx_ix_jdV=\frac{1}{c}\sum
(e_{ikl}x_j+e_{jkl}x_i)v_k\mu_l.
\label{eq20} \end{equation}

   The summing-up in (20) is done by different point magnetic
dipoles. By averaging over time and in view of the mean values being equal
to zero for the derivatives of the quantities varying within finite
limits with respect to time, we find:  \begin{equation}  \begin{array}{c}
\langle[\mb v\times\mb\mu]\rangle=-\langle[\mb r\times\dot{\mb\mu}]\rangle,
~~~\langle (e_{ikl}x_j+e_{jkl}x_i)v_k\mu_l\rangle= \\ \left\langle-
\frac12\left([\mb r\times\mb v]_i\mu_j+[\mb r\times\mb v]_j\mu_i\right)+
\delta_{ij}[\mb r\times\mb v]\cdot\mb\mu-\frac12 \left(x_i[\mb r\times
\dot{\mb\mu}]_j+x_j[\mb r\times\dot{\mb\mu}]_i\right)\right\rangle.
\end{array} \label{eq21} \end{equation}

   Using formulae (16),(20),(21) and dropping angular brackets, we find:
\begin{equation} \begin{array}{c}
\mb d^{(2)}=-\frac{1}{c}\sum[\mb r\times\dot{\mb\mu}], \\ Q^{(2)}_{ij}=
\frac{1}{2c}\sum\left(3[\mb r\times\mb v]_i\mu_j+3[\mb r\times\mb v]_j\mu_i
-2\delta_{ij}[\mb r\times\mb v]\cdot\mb\mu+3x_i[\mb r\times\dot{\mb\mu}]_j+
3x_j[\mb r\times\dot{\mb\mu}]_i\right), \\
\tau^{(2)}=\frac{2}{c}\sum([\mb r\times\mb v]\cdot\mb\mu).
\end{array} \label{eq22} \end{equation}

   By introducing the orbital moment $\mb l=[\mb r\times\mb p]=m_{rel}[\mb r
\times\mb v] ~~(m_{rel}=m/\sqrt{1-v^2/c^2}$ is the relativistic mass) we
transform (22) as follows:
\begin{equation} \begin{array}{c}
Q^{(2)}_{ij}=-\frac{1}{2m_{rel}c}\sum\left\{3l_i\mu_j+3l_j\mu_i-
2\delta_{ij}\mb l\cdot\mb\mu+3m_{rel}\left(x_i[\mb
r\times\dot{\mb\mu}]_j+x_j[\mb r\times\dot{\mb\mu}]_i\right)\right\}, \\
\tau^{(2)}=\frac{2}{m_{rel}c}\sum(\mb l\cdot\mb\mu).
\end{array} \label{eq23} \end{equation}

   The value $\dot{\mb\mu}$ in formulae (22),(23) for atoms
(nuclei) is determined by the electron (nucleon) spin precession. For the
states with the orbital moment $l\neq 0$, the terms containing
$\dot{\mb\mu}$ can be neglected [2,3]. In this case the formulae
for ECQM coincide with those given in these works. For
$l=0~(s$-states) the terms containing $\dot{\mb\mu}$ should be taken
into account in the calculations.

   The term proportional to $\dot{\mb\mu}$ in formula (23) for ECQM
agrees to the accuracy of the factor 3/2 with expression (12) for
the quadrupole moment created by the magnetic current $\mb
j^{(m)}=\sum_{i}\dot{\mb\mu}_i\delta(\mb r-\mb r_i)$. The presence of
this
additional factor in (23) is not surprising, because when averaging over
time as it has been done in the derivation of formulae (22),(23), we
determine the mean field at the fixed point of space. And in
this case $\left\langle\frac{\partial\mb A}{\partial t}\right\rangle=
\left\langle\frac{d\mb A}{dt}\right\rangle=0$, and there is no
contribution of the Miller multipoles to the mean field of the current
system. The Miller multipoles do not as well contribute to the charge
density the distribution of which determines
static quadrupole moments. However, we can say that the static ECQM
of the magnetic dipole system is the sum of the quadrupole
moments caused by the orbital motion of magnetic dipoles and
by the variation of the magnetic dipole moment values and their
orientation.

   The value $\tau$ defined by formula (23) describes the contribution
of the electrostatic current contact interaction (ECCI) to the value of
the total root-mean-square radius of the system. The ECCI does not
depend on $\dot{\mb\mu}$ in the system of magnetic dipoles. The classical
expression for the ECCI to the accuracy of the constant factor
corresponds to the quantum mechanical expression obtained in [4].

\begin{center}
{\bf  4. Conclusion} \end{center}

    The general formulae for the ECM appearing in the system of
conductors carrying a current or magnetic dipoles have been obtained in
terms  of  the  classical electrodynamics. There are electric current
multipole moments of the two types: 1) the moments
appearing at the translational motion of conductors (magnetic
multipoles); 2) the moments appearing at the current strength variation
in conductors or at the change of the magnetic dipole orientation. The
multipoles of the first type are created by electric charges appearing
at the motion of conductors carrying a current (magnetic multipoles). ECM
of this type appear both at the motion of particles having magnetic
multipole moments and in the rest system of compound
particles. The most important consequence of their existence is a
sufficiently great ECQM values for some nuclei in comparison with
their total quadrupole moments and the appearance
of ECQM for the moving anapole leading to the $P$-nonconservation at the
interaction of the particles having an anapole moment with
an external electrostatic field.  For the multipoles of the first
type electric charges are absent, and effective magnetic currents can be
introduced at the description of their field. For ECQM of
the both types the character of their motion in the electric field and
of the external charges in the ECM field is identical and
remains the same as for the electric charge
multipoles. At averaging the electric field over time
only the multipoles of the first type remain. The relationships obtained
for them can be transformed by introducing the orbital moment. In
particular, as a result, the dependence
of ECQM both on the orbital motion of magnetic dipoles and the
variation of their moment values or change of their spatial
orientation arises. The orbital motion of magnetic dipoles causes
the appearance of ECCI influencing the value of a charge
root-mean-square radius of the system.

\begin{center} {\bf References}
\end{center}

 1) J. Frenkel, Zeits. Phys. {\bf 37} (1926), 243.

 2) A.J. Silenko, Izv. VUZ Fiz. (Russian Phys. Rep.) {\bf 34} (1991),
No. 7, 53.

3) A.J. Silenko, J. of Phys. {\bf B25} (1992), 1661.

 4) A.J. Silenko, Izv. VUZ Fiz. (Russian Phys. Rep.) {\bf 38} (1995),
No. 4, 66.

     5) A.J. Silenko, Phys. Atom. Nucl. {\bf 60} (1997), 361.

 6) M.A. Miller, Usp. Fiz. Nauk (Sov. Phys.-Usp.) {\bf 142} (1984), 147.

   7) G.N. Afanasiev, Fiz. Elem. Chast. At. Yadra (Particl. and Nucl.)
{\bf 24} (1993), 512.

8) I.E. Tamm, {\it Electricity Theory Fundamentals} (Nauka, Moscow, 1989),
in Russian.

9) V.I. Strajev and L.M. Tomilchik, {\it Electrodynamics for
Magnetic Charge} (Nauka i technika, Minsk, 1975), in Russian.

     10) M.A. Miller, Radiophys. Quant. Electron. {\bf 29} (1986), 747.

     11) V.M. Dubovik, L.A. Tosunyan and V.V. Tugushev, Sov. Phys.-JETP
{\bf 63} (1986), 344.

     12) O.P. Sushkov, V.V. Flambaum and I.B. Khriplovich, Sov. Phys.-JETP
{\bf 60} (1985), 873.

   13) G.N. Afanasiev, Fiz. Elem. Chast. At. Yadra (Particl. and Nucl.)
{\bf 21} (1990), 172.

14) V.M. Dubovik, A.A. Cheshkov,  Fiz. Elem. Chast. At. Yadra
(Particl. and Nucl.) {\bf 5} (1974), 791.

15) M.C. Noecker, B.P. Masterson and C.E. Wieman, Phys. Rev.
Lett. {\bf 61} (1988), 310.

\end{document}